\documentclass[final,,numberedheadings,sort&compress]{aipproc}
\layoutstyle{6x9}
\usepackage{wrapfig}
\newcommand{\be}{\begin{equation}}
\newcommand{\ee}{\end{equation}}
\newcommand{\ba}{\begin{eqnarray}}
\newcommand{\ea}{\end{eqnarray}}
\newcommand{\la}{\langle}
\newcommand{\ra}{\rangle}
\newcommand{\di}{ {\rm d} }

\begin{document}
\title{The pretzelosity distribution function and intrinsic motion of the constituents in nucleon}

\classification{
      13.88.+e, 
      13.85.Ni, 
      13.60.-r, 
      13.85.Qk} 
\keywords{transverse momentum dependent distribution function,
      single spin asymmetry (SSA)}

\author{A.~V.~Efremov}{address={
  Bogoliubov Laboratory of Theoretical Physics, JINR, 141980 Dubna, Russia}}
\author{P.~Schweitzer}{address={
  Department of Physics, University of Connecticut, Storrs, CT 06269, USA}}
\author{O.~V.~Teryaev}{address={
  Bogoliubov Laboratory of Theoretical Physics, JINR, 141980 Dubna, Russia}}
\author{\underline{P.~Z\'{a}vada}}{address={
  Institute of Physics, Academy of Sciences of the Czech Republic,
  Na Slovance 2, CZ-182 21 Prague 8}}

\begin{abstract}
The pretzelosity distribution function $h_{1T}^\perp$ is studied 
in a covariant the quark-parton model which describes 
the structure of the nucleon in terms of 3D quark intrinsic motion. 
This relativistic model framework supports the relation between 
helicity, transversity and pretzelosity observed in other relativistic
models {\sl without} assuming SU(6) spin-flavor symmetry. 
Numerical results and predictions for SIDIS experiments are presented. 
\end{abstract}

\maketitle


\section{Introduction}
\label{Sec-1:introduction}

Transverse parton momentum dependent parton distribution (TMDs) and 
fragmentation functions \cite{Collins:2003fm,Collins:1992kk,Mulders:1995dh,Boer:1997nt,Brodsky:2002cx,Collins:2002kn}
offer the access to novel information on the nucleon structure
\cite{Miller:2007ae}.
TMDs can be accessed in processes like semi-inclusive 
deep-inelastic lepton nucleon scattering (SIDIS) 
\cite{Ji:2004wu}.
Data on such reactions 
\cite{Arneodo:1986cf,Airapetian:1999tv,Avakian:2003pk,Airapetian:2004tw,Alexakhin:2005iw,Kotzinian:2007uv}
provide first insights
\cite{Efremov:2002ut,Efremov:2004tp,Vogelsang:2005cs,Efremov:2006qm,Anselmino:2007fs,Arnold:2008ap,Zhang:2008nu}.
However, model calculations play an important role for the understanding of
the novel functions 
\cite{Jakob:1997wg,Avakian:2008dz,Pasquini:2008ax,Bacchetta:2008af,Meissner:2007rx,Yuan:2003wk,Gamberg:2007gb,Schweitzer:2001sr,Pasquini:2005dk,Zavada:1996kp,Zavada:2001bq,Efremov:2004tz,Zavada:2006yz}.

An important question in this context is whether it is possible to
relate unknown TMDs with possibly better known ones. Such relations
cannot be exact, since all TMDs are independent. 
Approximations motivated partly by data were discussed in 
\cite{Avakian:2007mv}. The ideal playground to motivate and test 
any such relations among TMDs are models.

An interesting relation between pretzelosity $h_{1T}^{\perp q}$, 
transversity $h_1^q$ and helicity~$g_1^q$ was observed in bag model 
\cite{Avakian:2008dz}. The name pretzelosity reflects that this function 
'measures' an appropriately defined deviation of the nucleon from spherical 
shape which could look like a pretzel \cite{Miller:2007ae}.
This relation holds also in the spectator model \cite{Jakob:1997wg}, and was 
subsequently confirmed in the constituent quark model \cite{Pasquini:2008ax}
but not in the model of \cite{Bacchetta:2008af}.

The purpose of this work is to study pretzelosity, and its possible
relations, in the covariant model of the nucleon of Ref.~\cite{Zavada:1996kp}. 
In this model the intrinsic motion of partons inside the nucleon is described 
in terms of a covariant momentum distribution. The model was applied 
to the study of unpolarized and polarized structure functions 
accessible in DIS $f_1^a(x)$, $g_1^a(x)$ and $g_T^a(x)$
\cite{Zavada:1996kp,Zavada:2001bq} and extended to compute transversity  
$h_1^a(x)$ \cite{Efremov:2004tz}.
In this work we will generalize the approach to the description of TMDs,
focusing on chiral-odd TMDs accessible with transverse nucleon polarization.

\section{Chiral-odd TMDs with transverse polarization}
\label{Sec-2:TMDs}

We focus on chiral-odd TMDs in a nucleon polarized transversely,
e.g.\ in SIDIS, with respect to the hard virtual photon 
$q^\mu=(q^0,|\vec{q}|,0,0)$. 
The light-front quark-correlator with the process-dependent 
Wilson-link ${\cal W}$ \cite{Collins:2002kn}
where $z^\pm=(z^0\pm z^1)/\sqrt{2}$ etc.,
\be\label{Eq:correlator}
    \phi(x,\vec{p}_T,\vec{S}_T)_{ij}=
    \int\frac{\di z^-\di^2\vec{z}_T}{(2\pi)^3}\;e^{ipz} \; 
    \la P,\vec{S}_T|\bar\psi_j(0)\,{\cal W}(0,z,\mbox{path})\,\psi_i(z)
    |P,\vec{S}_T\ra \biggl|_{z^+=0,\,p^+ = xP^+} \;,
    \ee
allows to define (3 out of the 4) chiral-odd TMDs in the nucleon as follows
\ba\label{Eq:TMD-pdfs-III}
    \frac12\;{\rm tr}\biggl[i\sigma^{j+}\gamma_5 \;
    \phi(x,\vec{p}_T,\vec{S}_T)\biggr] &=&
    S_T^j\,h_1  + 
    \frac{(p_T^j p_T^k-\frac12\,\vec{p}_T^{\:2}\delta^{jk})S_T^k}{M_N^2}\,
    h_{1T}^\perp + \frac{\varepsilon^{jk}p_T^k}{M_N}\,h_1^\perp \;,
\ea
where $\varepsilon^{32} = - \varepsilon^{23} = 1$ 
and zero else.
The only structure surviving the $\vec{p}_T$-integration 
in (\ref{Eq:correlator}) is transversity $h_1^a(x)$. Nucleon polarizations and 
Dirac-structures other than that in
Eqs.~(\ref{Eq:correlator},~\ref{Eq:TMD-pdfs-III})
lead to further leading- and subleading-twist TMDs
\cite{Mulders:1995dh,Boer:1997nt,Bacchetta:2006tn}.

\section{The covariant model of the nucleon
and TMDs}
\label{Sec-3:model}

The starting point for the calculation of the chiral-even 
functions accessible in DIS, $f_1^a(x)$, $g_1^a(x)$, $g_T^a(x)$, 
is the hadronic tensor \cite{Zavada:1996kp,Zavada:2001bq}.
In the model it is assumed that DIS can be described as the incoherent 
sum of the scattering of electrons off non-interacting quarks, whose 
momentum distributions inside the nucleon are given in terms of the 
scalar functions: $G=G^\uparrow+G^\downarrow$ for unpolarized
and $H=G^\uparrow-G^\downarrow$ for polarized quarks. 

$G^i(pP/M)$ denotes the distribution of quarks of some (not indicated)
flavour that are polarized parallel (antiparallel) $\uparrow$($\downarrow$)
to the $i$-axis, where $p$ is the quark momentum and $M$ the nucleon mass.
Though all expressions can be formulated in a manifestly
covariant way, it is convenient to work in the nucleon rest-frame,
where the $G^i$ become functions of $p^0=\sqrt{\vec{p}^{\;2}+m^2}$ 
with $m$ the quark mass, and the distributions are rotationally symmetric.

The chiral-odd $h_1^q(x)$ cannot be accessed in DIS through the hadronic 
tensor. However, for theoretical purposes one may consider the auxiliary 
process described by the interference of a vector and a scalar current,
described on the quark level  by
$T_\alpha^q=\epsilon_{\alpha\beta\lambda\nu} p^\beta q^\lambda w^\nu$
where $w^\nu$ is the quark polarization vector.
The nucleon current follows from convoluting $T_\alpha^q$ with the 
momentum distribution of polarized quarks $H(p^0)$ and reads
\be\label{Eq:aux-vector}
        T_\alpha(x) = 
         \frac{1}{2Pq}\epsilon_{\alpha\beta\lambda\nu}q^\lambda
         \int\frac{\di^3 p}{p^0}\;H(p^0)
         \delta\biggl(\frac{p^0-p^1}{M}-x\biggr) p^\beta w^\nu \;.
\ee
The auxiliary current is related to transversity as
\be\label{Eq:aux-vector-gen}
         2M\,T_\alpha(x)\,\epsilon^{\alpha j} = S_T^j \,h_1^q(x)\;.
\ee

Before attempting to extend the approach to TMDs, 
let us stress that the QCD definition of TMDs
includes a Wilson line absent in our model with no gauge boson
degrees of freedom. In such an approach time-reversal (T) odd TMDs, 
such as the Boer-Mulders function $h_1^\perp$ in (\ref{Eq:TMD-pdfs-III}), 
vanish \cite{Brodsky:2002cx,Collins:2002kn}.

Now we turn to the question how to extend the approach to describe
of TMDs, focusing here on chiral-odd ones in a transversely polarized 
nucleon. For that we observe that the expression for the auxiliary current 
(\ref{Eq:aux-vector}) is of the type:
$T_\alpha(x) = \int\di^2 p_T T_\alpha(x,\vec{p}_T)$.
In the following we explore the consequences of what happens if one 
{\sl does not integrate out} transverse momenta in this expression.

With $S^\mu$ denoting the nucleon polarization vector 
(here $S^\mu=(0,0,\vec{S}_T)$ with $|\vec{S}_T|=1$) 
the most general expression \cite{Zavada:2001bq} for the 
covariant quark polarization vector $w^\mu$ reads
\be\label{Eq:quark-pol-vec}
         w^\mu = - \frac{pS}{pP+mM}\,P^\mu + S^\mu 
                 - \frac{M}{m}\,\frac{pS}{pP+mM}\,p^\mu \;.
\ee
From (\ref{Eq:quark-pol-vec}) we obtain for the 'unintegrated' 
auxiliary current contracted with $\varepsilon^{\alpha j}$
the result
\be
	2M\,T_\alpha(x,\vec{p}_T) \,\varepsilon^{\alpha j}
	= 
	\int\frac{\di p^1}{p^0}\;H(p^0)\;\delta\left(\frac{p^0-p^1}{M}-x\right)
	\Biggl\{S_T^j (p^0-p^1) - p_T^j\;\frac{\vec{S}_T\vec{p}_T}{p^0+m}
	\Biggr\} \, .\label{Eq:aux-vector-unintegrated} \ee
By comparing to (\ref{Eq:TMD-pdfs-III}) we read off the following results:
\ba
	h_1^q(x,p_T) &=& 
	\int\frac{\di p^1}{p^0}\;H(p^0)\;\delta\left(\frac{p^0-p^1}{M}-x\right)
	\biggl[p^0-p^1-\frac{\vec{p}_T^{\:2}}{2(p^0+m)}\biggr],\label{Eq:h1}\\
	h_{1T}^{\perp q}(x,p_T) &=& 
	\int\frac{\di p^1}{p^0}\;H(p^0)\;\delta\left(\frac{p^0-p^1}{M}-x\right)
	\biggl[-\frac{M^2}{p^0+m}\biggr],\label{Eq:pretzelosity}
\ea
and $h_1^{\perp q}(x,p_T)=0$. Several comments are in order. First, 
in our approach the vanishing of the T-odd $h_1^{\perp q}$ is consistent.
Second, integrating in Eq.~(\ref{Eq:h1}) over $\vec{p}_T$ yields the model 
expression for $h_1^q(x)\equiv\delta q(x)$ from \cite{Efremov:2004tz}. 
Third, $h_{1T}^{\perp q}\neq 0$ implies non-sphericity in the nucleon
in the sense of \cite{Miller:2007ae} inspite of a spherically symmetric 
$H(p_0)$. Forth, adding $h_1^q(x)$ and $h_{1T}^{\perp(1)q}(x)=$
$\int\di^2p_T\frac{\vec{p}_T^{\:2}}{2M^2}\,h_{1T}^{\perp q}(x,p_T)$ 
yields the model expression for $g_1^q(x)\equiv\Delta q(x)$ 
derived in \cite{Zavada:2001bq}, i.e.\ we recover the remarkable relation 
\cite{Avakian:2008dz}:
\be\label{Eq:relation}
        g_1^q(x) - h_1^q(x) = h_{1T}^{\perp(1)q}(x) \;.
\ee
This relation is satisfied in several 
\cite{Jakob:1997wg,Avakian:2008dz,Pasquini:2008ax} though 
not all \cite{Bacchetta:2008af} quark models.
Remarkably, it follows in our approach 
{\sl without} assuming SU(6) spin-flavour symmetry of the nucleon wave 
function as was done in \cite{Jakob:1997wg,Avakian:2008dz,Pasquini:2008ax}.
This is an important observation: SU(6) is not a necessary condition for 
the relation (\ref{Eq:relation}) to be satisfied in a {\sl quark model}.
What is a necessary condition is the absence of gluon degrees of freedom
\cite{Meissner:2007rx}.

Finally, we remark that in the chiral limit $m\to 0$ it is possible to 
relate the transverse moment of pretzelosity to the twist-3 parton 
distribution function $g_T^q(x)$ \cite{Zavada:2001bq} as follows
\be\label{Eq:rel-chi-lim-4}
        h_{1T}^{\perp(1)q}(x) + g_T^q(x) = {\cal O}\biggl(\frac{m}{M}\biggr)\;.
\ee
Since in the model the WW-relation holds,
$g_T^q(x)=\int_x^1\di y \,g_1^q (y)/y + {\cal O}(\frac{m}{M})$
\cite{Zavada:2001bq}, this offers a possibility to 
{\sl estimate} pretzelosity numerically in the model framework.

\section{Results and phenomenology}

We estimate $h_{1T}^{\perp(1)q}(x)$ in our approach using
(\ref{Eq:rel-chi-lim-4}) and the WW-approximation for $g_T^q(x)$
with $g_1^q(x)$ at a scale of $2.5\,{\rm GeV}^2$ from \cite{Gluck:2000dy}.
We obtain the results shown in Fig.$\,1$a.

The azimuthal SSA from transversely polarized targets,
$A_{UT}^{\sin(3\phi-\phi_S)}\propto\sum_ae_a^2h_{1T}^{\perp(1)a}H_1^{\perp a}$,
allows to access pretzelosity in SIDIS due to the Collins effect 
\cite{Collins:1992kk}, see \cite{Mulders:1995dh,Avakian:2008dz} 
for details. We use the information on $H_1^\perp$ from 
\cite{Vogelsang:2005cs,Efremov:2006qm,Anselmino:2007fs}.
Fig.$\,1$b shows that the model results for the SSA are compatible 
with preliminary COMPASS deuteron target data \cite{Kotzinian:2007uv}.
Fig.$\,1$c shows estimates for the SSA in the kinematics of the CLAS 
$12\,{\rm GeV}$ beam experiment. The error projections from 
\cite{Avakian-LOI-CLAS12} included in the plot demonstrate that CLAS 
will be able to measure effects of pretzelosity of the size predicted 
by the model.

\begin{figure}[t!]
\begin{tabular}{ccc}
\includegraphics[height=5cm]{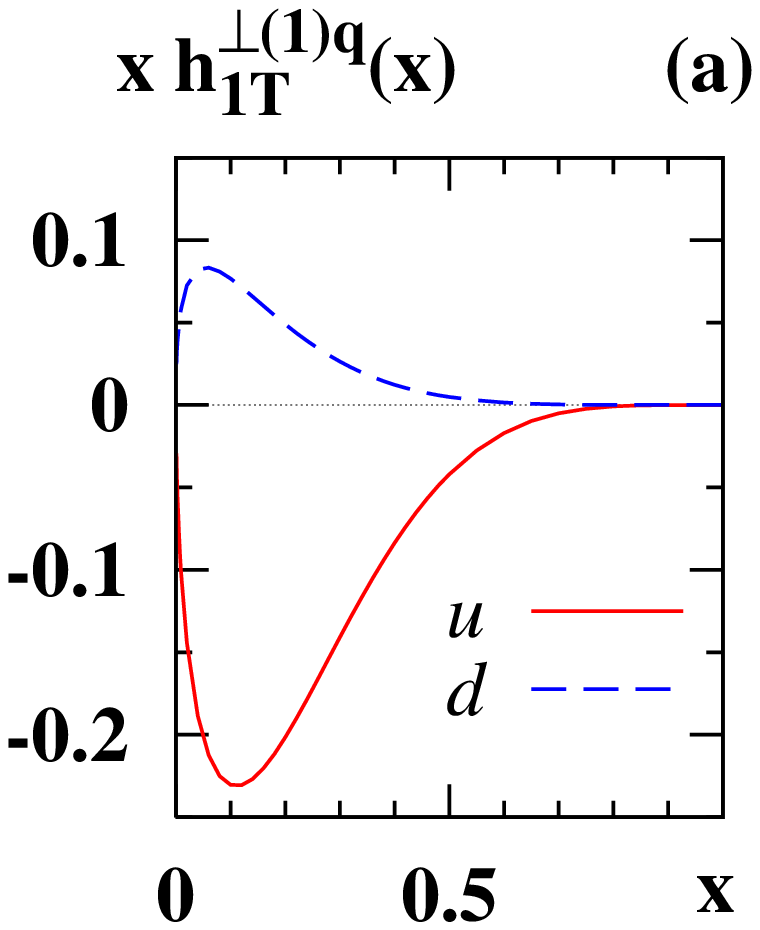}&
\includegraphics[height=5cm]{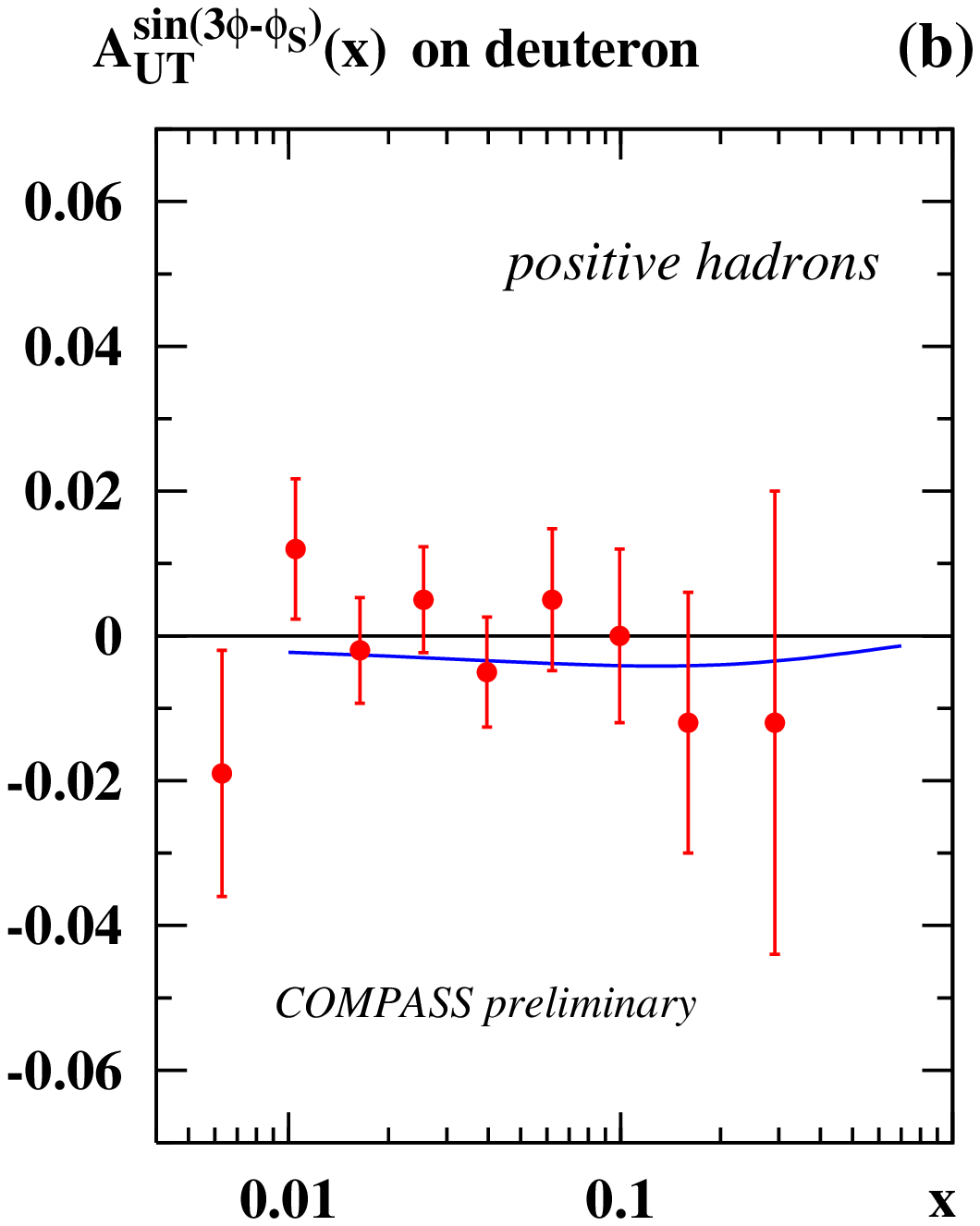} &
\includegraphics[height=5cm]{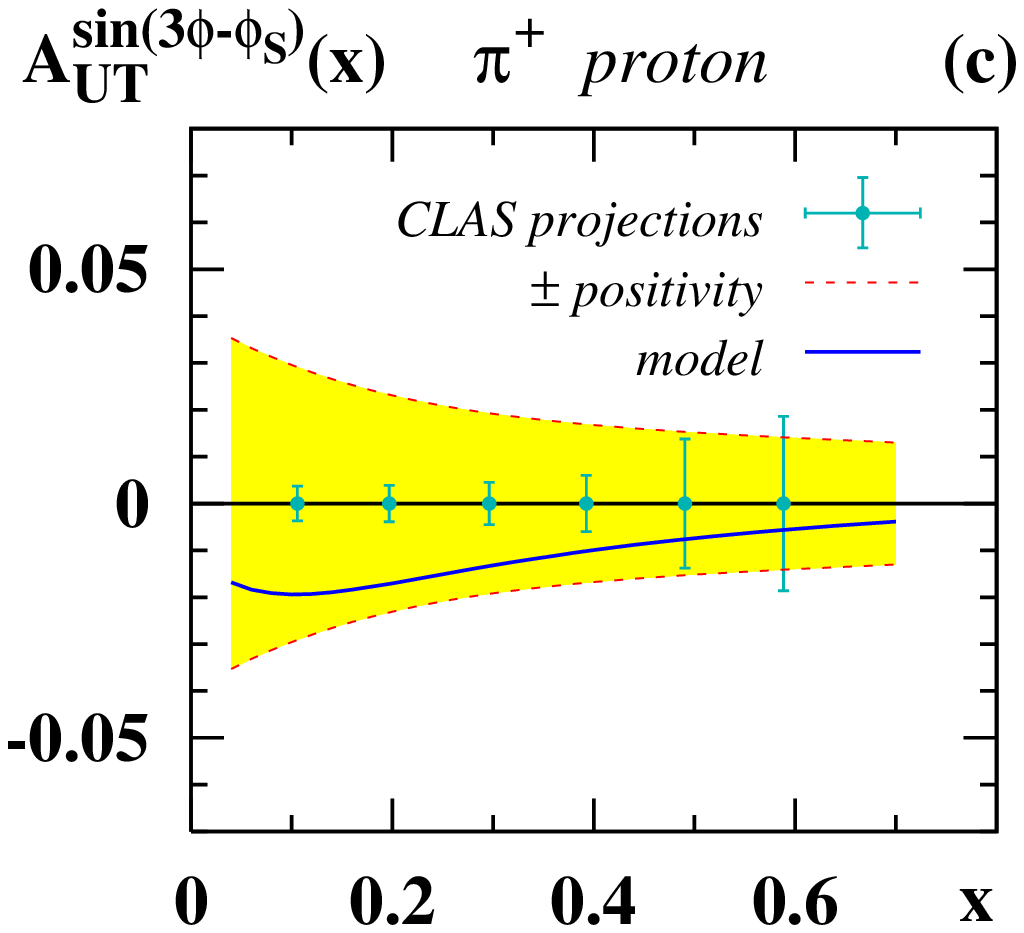} 
\end{tabular}
     \caption{\label{Fig1}
     (a) $x\,h_{1T}^{\perp(1)q}(x)$ vs.\ $x$ from the present approach.
     (b) $A_{UT}^{\sin(3\phi-\phi_S)}$ as function of $x$ computed with 
     pretzelosity from Fig.~1a for positive hadrons from a deuterium target 
     at COMPASS in comparison to preliminary data \cite{Kotzinian:2007uv}.
     (c) The same as Fig.~1b but for $\pi^+$ from proton target at CLAS.
     The shaded area is the region allowed by positivity \cite{Avakian:2008dz},
     the error projections are from \cite{Avakian-LOI-CLAS12}.}
\end{figure}

\section{Conclusions}

A generalization of the covariant model 
\cite{Zavada:1996kp,Zavada:2001bq,Efremov:2004tz} to the description of 
TMDs was suggested, and applied to compute the pretzelosity distribution 
function $h_{1T}^\perp$. In particular, it was shown that the relation 
between helicity, transversity and pretzelosity \cite{Avakian:2008dz} 
is satisfied in this model --- remarkably, without assuming SU(6) symmetry.

A numerical estimate of $h_{1T}^{\perp(1)q}$ was presented, and 
used to compute $A_{UT}^{\sin(3\phi-\phi_S)}$, the leading-twist SSA
in SIDIS due to Collins effect and pretetzelosity.
The model results are compatible with the preliminary deuteron
target data from COMPASS \cite{Kotzinian:2007uv}. 
Predictions of this observable in the
kinematics of the CLAS experiment with upgraded $12\,{\rm GeV}$ beam
suggest that information on pretzelosity is accessible at
Jefferson Lab \cite{Avakian-LOI-CLAS12}.

\begin{theacknowledgments}
A.~E.\ and O.~T.\ are supported by the Grants RFBR 06-02-16215 and
07-02-91557, RF MSE RNP.2.2.2.2.6546 (MIREA) and by the Heisenberg-Landau
and (also P.Z.) Votruba-Blokhitsev Programs of JINR.
P.~Z.\ is supported by the project AV0Z10100502 of the Academy of
Sciences of the Czech Republic.
\end{theacknowledgments}

\end{document}